\documentclass[11pt]{article}

\usepackage{amssymb}
\usepackage{epsfig}
\usepackage[dvips]{color}
\usepackage[authoryear]{natbib}

\newcommand{\eg}{e.\,g.}
\newcommand{\ie}{i.\,e.}
\newcommand{\wrt}{w.\,r.\,t.}

\newcommand{\GOI}{{\sc goi}}
\newcommand{\LI}{{\sc li}}
\newcommand{\PDF}{{\sc pdf}}
\newcommand{\PM}{{\sc pm}}
\newcommand{\RO}{{\sc ro}}
\newcommand{\VHI}{{\sc vhi}}

\newcommand{\beq}{\begin{equation}}
\newcommand{\eeq}{\end{equation}}
\newcommand{\beqa}{\begin{eqnarray}}
\newcommand{\eeqa}{\end{eqnarray}}
\newcommand{\beqas}{\begin{eqnarray*}}
\newcommand{\eeqas}{\end{eqnarray*}}
\newcommand{\ceq}{\!\!\! & = & \!\!\!}
\newcommand{\real}{{\ensuremath{\mathbb{R}}}}
\newcommand{\dx}{{\dot x}}
\newcommand{\ddx}{{\ddot x}}
\newcommand{\pd}{{\partial}}

\newcommand{\ts}{{\widetilde\sigma}}
\newcommand{\wtt}{{\widetilde\tau}}
\newcommand{\wht}{{\widehat\tau}}

\newcommand{\Ha}{{H_\alpha}}
\newcommand{\hp}{\mathcal{H}}
\newcommand{\X}{\mathcal{X}}
\newcommand{\Fe}{\mathfrak{e}}
\newcommand{\mfour}[4]{%				% 2 x 2 matrix
\left(\begin{array}{cc}
\scriptstyle{#1}&\scriptstyle{#2}\\\scriptstyle{#3}&\scriptstyle{#4}
\end{array}\right)}

\newcommand{\mytitle}[3]{%
\begin{center}
\LARGE #1
\par\bigskip
\normalsize\rm #2
\par\medskip
\small #3
\end{center}
}

\newcommand{\putfig}[2]{\epsfig{file=#1.eps,scale=#2}}

\newcommand{\sep}{, }

\newcommand{\book}[5]{{#1} (#5). {\it #2}. {#4}: {#3}.}
\newcommand{\jour}[6]{{#1} (#6). {#2} {\it #3}, {\it #4}, {#5}.}
\newcommand{\prop}[8]{%
{#1} (#8). {#2} In #3 (Eds.), {\it #4} (p.~#7). {#6}: {#5}.}
\newcommand{\proc}[8]{%
{#1} (#8). {#2} In #3 (Eds.), {\it #4} (pp.~#7). {#6}: {#5}.}

\begin{document}

\mytitle{Geometric--optical illusions and Riemannian geometry}
{Werner Ehm$^{a}$ and Ji\v{r}\'i Wackermann$^{b}$}
{\footnotesize{$^{a}$ Heidelberg Institute for Theoretical Studies, Schloss-Wolfsbrunnenweg 35, 69118 Heidelberg, Germany, {\tt werner.ehm@h-its.org}}}

\vspace{-4mm}
\centerline{{\footnotesize{$^{b}$ Independent researcher, 79261 Gutach, Germany, {\tt mail@jiri-wackermann.eu}}}}

\smallskip
\begin{abstract}
\noindent
Geometric-optical illusions (\GOI) are a subclass of a vast variety of visual illusions. A special class of {\GOI}s originates from the superposition of a simple geometric figure (``target'') with an array of non-intersecting curvilinear elements (``context'') that elicits a perceptual distortion of the target element. Here we specifically deal with the case of circular targets. Starting from the fact that (half)circles are geodesics in a model of hyperbolic geometry, we conceive of the deformations of the target as resulting from a context-induced perturbation of that ``base'' geometry. We present computational methods for predicting distorted shapes of the target in different contexts, and we report the results of a psychophysical pilot experiment with eight subjects and four contexts to test the predictions. Finally, we propose a common scheme for modeling {\GOI}s associated with more general types of target curves, subsuming those studied previously.

\smallskip
\noindent
{\em Keywords}:  calculus of variations\sep
  Ehrenstein--Orbison type illusions\sep
  geodesic\sep
  geo\-met\-ric--optical illusions\sep
  Poincar\'e model\sep
  Riemannian geometry\sep
  vector field\sep
  visual perception

\smallskip
\noindent
{\em Abbreviations}:  
\GOI: geometric--optical illusion\sep
\LI: local interactions\sep
\PDF: Portable Document Format\sep
\PM: Poincar\'e model\sep
\RO: regression to orthogonality\sep
\VHI: vertical--horizontal illusion
\end{abstract}

%\smallskip
\normalsize
% ----------------------------------------------------------------------
\section{Introduction}

Visual perception informs us about the outward reality in the surrounding space. Under certain circumstances, the result of a perceptual process (``percept'', for short) may remarkably differ from our knowledge of the objective reality as it is evidenced, \eg, by measurement results, cognitive inferences, or other percepts. Those situations are commonly known as ``visual illusions.''\footnote{%
Of course,  perceptual illusions are known also in other sensory modalities, or occurring as inter-modal interactions. However, these types of illusions are not in our focus, nor are visual illusions affecting optical qualities such as brightness or color, and phenomena involving illusory motion. An exhaustive overview of the field is beyond the scope of the present paper.}
Visual illusions are not deliberate deceptions or random errors of the visual system; they are systematically occurring, experimentally reproducible and measurable phenomena, presumably revealing essential properties of the visual system, and as such they are a proper subject of scientific study \citep{Met1975,CoGi1978,Rob1998,Eag2001}. Following Gregory's (\citeyear{Gre1997b}) proposal, visual illusions can be roughly subdivided into four types: fictions, paradoxes, ambiguities, and distortions.

Geometric--optical illusions (\GOI) are an interesting subclass of visual {\em distortions}. In {\GOI}s, geometric properties of a stimulus---\eg, lengths, angles, areas, or forms---are affected and systematically altered by the presence of other elements in the visual field. For example, a straight line appears slightly curved when superposed with an array of straight or curved lines \citep{Her1861}; a~length marked by two distinct elements appears larger if the space between them is subdivided by additional elements \citep{Opp1861,Kun1863}; two equally long line segments appear different when marked by arrows of opposite orientation \citep[probably the most popular \GOI]{ML1889}; etc.  These phenomena were discovered and named about one and a half century ago \citep{Opp1855}, and since then their number significantly expanded. Despite numerous classificatory \citep{Cor1976,Gre1997b} and explanatory attempts \citep{Cha2008, CoGi1978} based on optical, retinal, cortical or cognitive mechanisms, there is by now no unitary theory of {\GOI}s, let alone of visual illusions on the whole. There is not even consensus about general principles upon which such a theory could or should be based \citep{Wac2010,Zav2015}. Nonetheless, the {\GOI}s deserve special attention: not only because, historically, they ``form the core of the subject [of visual illusions]'' \citep[p.~11]{Rob1998}, but also for their link to geometry-based theories of visual perception.

In the present paper we aim at a {\em phenomenological} theory of {\GOI}s; that is, we search for a mathematical representation of the phenomena under study, not for an explanation via physiological or psychological mechanisms. Our focus is on a special class of {\GOI}s based on interactions between a {\em target} element and {\em context} elements in the visual field. In our previous work \citep{EW2012} we were studying Hering type illusions, where the target was a segment of a straight line. Here we consider the case where the target is a {\em circle}, such as in Figs.~1a, 1b. These and similar illusions were described, independently and in different conceptual frameworks, by \citet{Ehr1925} and \citet{Orb1939}. Illusions of this type and of Hering type have in common the {\em angular expansion} effect, also known as ``regression to right angles'' \citep{HoOl1972,HoRo1975}: the illusory distortion of the target acts so as to enlarge the acute angles at the intersection points. It is thus plausible to assume that a common approach may account for both groups of phenomena.

\begin{figure}[ht]	%%%%%%%	Figure 1 %%%%%%%
\begin{center}
\putfig{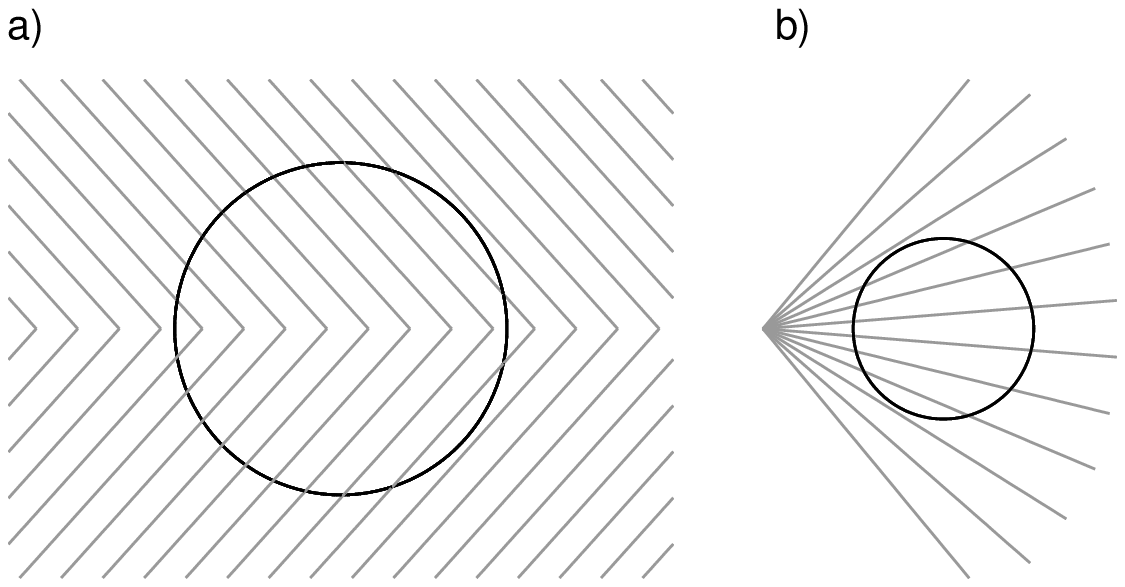}{1}
\end{center}
\small
Figure~1. Two examples of \GOI{s} with a circular target, adapted from \citet{Orb1939} and \citet{Ehr1925}. Although the targets are perfect circles, they appear to the observer as being slightly inward-dented on the right-hand side (a) or on the left-hand side (b).
\end{figure}

In our earlier paper \citep{EW2012} we modeled the distorted percepts of a straight line segment (target) by the solutions of a variational problem, namely as the shortest path connecting the endpoints of the target when length is measured in terms of a context-induced perturbation of the Euclidean metric. In regard to the circular targets in Ehrenstein--Orbison type illusions the questions arise: can this principle be generalized? And, if so, how to overcome the restriction to straight-line targets inherent in the original approach? The basic idea permitting an extension to curved targets consists in {\em equipping the target itself with a geometry} \citep{EW2013}. In this view both the target and the distorted percepts figure as paths of shortest length (geodesics) in an appropriate ``base'' geometry and a context-induced perturbation of that geometry, respectively. In the present paper we elaborate on this approach, using the fact that half circles represent geodesics in a suitable model of hyperbolic geometry.

The elements of this framework are presented in Section~2. We introduce the base and the perturbed hyperbolic geometries and describe the single steps leading to our final prediction of the distorted percept. The details are deferred to the mathematical appendices. An experiment intended to verify the predictions and to measure the magnitude of the illusory distortion is reported in Section~3. The discussion in the final Section~4 addresses phenomenological as well as modeling aspects. It concludes with the above indicated proposal for a general approach covering \GOI{s} of the Hering and the Ehrenstein--Orbison type as special cases.

% ----------------------------------------------------------------------
\section{Mathematical model}

\subsection{Preliminary remarks}

Our mathematical description of the visual distortion of the target figure draws on minimum principles related to geometrical conceptions. We think of the context figure as distorting the spatial relationships between the points of the drawing plane, similar to when an elastic substance is kneaded: some portions are expanded while others are condensed. Riemannian geometry \citep{Lau1965} makes it possible to describe such situations mathematically by means of a locally varying metric that attaches a well-defined length to every path through the respective range. The connection to minimum principles comes via the concept of a {\em geodesic,} which is a path of minimal length, measured in the respective metric, that connects two given points. In fact, geodesics also represent paths requiring minimum energy or effort, rendering them relevant to fundamental conceptions about human perception; cf.~Section \ref{explstatus}.

According to our basic surmise the percept of the circular target can be modeled as a geodesic in a suitable geometry that is perturbed by the context if such is present. It is known that a (2D) Riemannian geometry admitting only circles and straight lines as geodesics must have constant curvature (Khovanskii, 1980). The possible candidates for a ``suitable'' or {\em base} geometry thus must be one of the classical elliptic, Euclidean, and hyperbolic geometries. Among these only Poincar\'e's half plane model of hyperbolic geometry [\PM] fulfills the following three conditions essential to our approach.\footnote{%
Strictly speaking, Poincar\'e's disc model \citep{Can1997} satifies (i) to (iii), too. However, it serves our purposes less well as the geodesics span less than a half circle, and the contexts have to be placed in the disc; both circumstances imply substantial restrictions.}
\begin{enumerate}
\item[i)]
The base geometry has to have circle segments as geodesics---so does the \PM.
\item[ii)]
The perceptual distortions are hypothesized to obey the local interactions principle, particulary, to depend on the intersection angles between the context and the target lines. Therefore, the intersection angles should be represented faithfully---the \PM\ is conformal.
\item[iii)]
The relevance of geometries derived from immersions of a surface into $\real^3$ appears doubtful considering that our stimulus figures are presented, and seen, on a flat screen---the \PM\ implements the distortions of the Euclidean metric entirely within (a subset of) $\real^2$.
\end{enumerate}

To summarize, adopting the \PM\ for the base geometry is virtually cogent for our modeling approach. The \PM\ allows us to deal with circular targets essentially along the same lines as with the straight line targets considered in the earlier work (\citeyear{EW2012}), although the details are rather more involved. In this section we only present the main lines of our geometrical approach; the mathematical elaboration is postponed to the appendices.

\subsection{Poincar\'e model of the hyperbolic plane}

The Poincar\'e model equips the upper half plane $\hp = \{\xi = (\xi_1,\xi_2) \in \real^2,\, \xi_2 > 0\}$ with the line element $ds^2 = (d\xi_1^2 + d\xi_2^2)/\xi_2^2$. (For comparison, the Euclidean line element is $d\xi_1^2 + d\xi_2^2$.) This means that the length of a smooth curve $t \mapsto x(t) = (x_1(t),x_2(t)),\, t_0\leq t \leq t_1$ in $\hp$ is, invariantly under reparameterization, defined as 
\beq\label{length}
L(x) = \int_{t_0}^{t_1} \frac{|\dx(t)|}{x_2(t)}\, dt = \int_{t_0}^{t_1} \sqrt{\langle \dx(t), H(x(t))\, \dx(t) \rangle}\, dt.
\eeq
Here $\dx$ stands for $dx/dt = (dx_1/dt,dx_2/dt)$, and $| \cdot|$ and $\langle \cdot, \cdot \rangle$ denote the standard Euclidean norm and inner product, respectively. In the last term of (\ref{length}), 
\beq\label{metric}
H(\xi) = \xi_2^{-2} I\quad (\xi \in \hp) \qquad \mbox{where} \qquad I = \mfour{1}{0}{0}{1}.
\eeq
The matrices $H(\xi)$ give rise to the metric tensor of the Riemannian geometry associated with the Poincar\'e model. Since they are scalar multiples of the identity matrix $I$ it is evident that angles are preserved in this geometry, that is, they are identical with the Euclidean angles. Lengths, however, are not preserved, and geodesics need not be ``straight'' in the usual sense. In fact, all geodesics in the hyperbolic plane are segments of the following two types of curves: (a)~vertical lines in $\hp$ (parallel to the $x_2$ axis); (b)~half circles in $\hp$ orthogonal to the $x_1$ axis (satisfying $(x_1-a)^2 + x_2^2 = r^2$ for some $a \in \real,\,r >0$); see Fig.~2.

\begin{figure}[ht]	%%%%%%%	Figure 2 %%%%%%%
\putfig{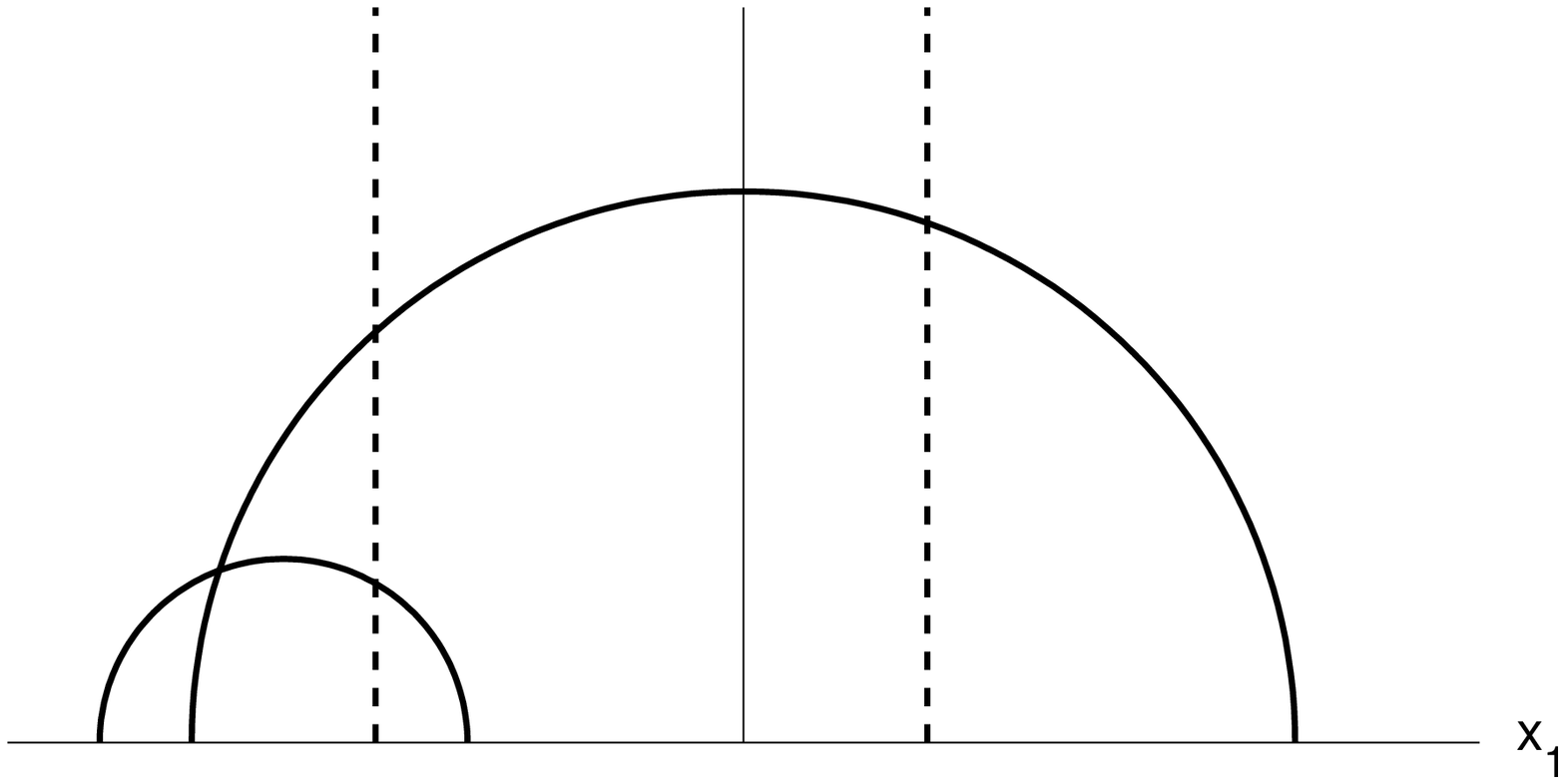}{0.4}%
\hspace{2mm}
\parbox[b]{7.4cm}{
\footnotesize
Figure~2. Geodesics in the Poincar\'e model: half circles and straight lines orthogonal to the $x_1$ axis.
\vspace{18ex}
}
\end{figure}

\subsection{Target, context, and the perturbed hyperbolic geometry}
\label{taco}

Circles with center at the origin will serve as the target component of our stimulus figures. The context component is represented by means of a smooth planar vector field of unit directions $v(\xi)\ (|v(\xi)| = 1, \, \xi \in \Xi;\ v$ twice continuously differentiable) defined on a region $\Xi$ that contains the target. A finite sample of the stream lines of $v$ makes up the context curves actually presented to the observer. 

The Poincar\'e model requires decomposing the complete figure into the two parts contained in the upper and the lower half planes, respectively. These two parts will be treated separately. At first, we consider the upper part, and (segments of) upper half circles as targets. 
Given a parameter $\alpha \ge 0$ that accounts for the strength of the distortion, the context-pertubed Riemannian geometry in the upper half plane is determined by declaring the length of a curve $x$ in $\hp$ as
\beq\label{lengthp}
L_\alpha(x) = \int_{t_0}^{t_1} \sqrt{\langle \dx(t), \Ha(x(t))\, \dx(t) \rangle}\, dt, 
\eeq
with positive definite matrices 
\beq\label{pmetric}
\Ha(\xi) = \xi_2^{-2} G_\alpha(\xi) \quad \mbox{where} \quad G_\alpha(\xi) = I + 2 \alpha\, v(\xi)\!\otimes\! v(\xi) \qquad (\xi \in \Xi \cap \hp).
\eeq
This definition implements two basic principles already employed by \citet{EW2012}:

%\vspace*{-1ex}
\begin{enumerate}
\setlength{\itemsep}{0pt}
\item[(a)]
the {\em local interactions} [\LI] hypothesis: the context $v$ ``acts'' only along the candidate curves, in the vicinity of the target;

\item[(b)]
the {\em regression to orthogonality} [\RO] hypothesis: the distortion operates toward an expansion of the acute angles at the points of intersection.
\end{enumerate}
Both these hypotheses, supported by experimental phenomenology \citep[cf.][]{Hor1971,Wac2010}, are built into the expression for the curve length $L_\alpha(x)$, which when evaluated explicitly becomes
\beq\label{lenxpl}
L_\alpha(x) = \int_{t_0}^{t_1} \frac{\sqrt{|\dx(t)|^2 + 2\alpha \langle \dx(t), v(x(t)) \rangle^2}}{x_2(t)}\, dt.
\eeq
By (\ref{lenxpl}), the context vector field enters $L_\alpha(x)$ only locally at the curve $x$ [\LI]; and the term $\langle \dx(t), v(x(t)) \rangle^2$ penalizes non-orthogonality (for positive $\alpha$) between the vector field $v$ and the tangents of $x$ [\RO]. A geodesic thus realizes an optimal compromise between being {\em as short as possible}, and passing through the stream lines of $v$ {\em as orthogonally as possible}. We denote the $\Ha$ geodesic with the same endpoints as the target as $\gamma_\alpha$. It represents our ideal candidate for the prediction of the actual, distorted percept of the target stimulus. A necessary condition for a curve $x$ to be a geodesic in the $\Ha$ metric is given in the form of a second-order nonlinear differential equation (Appendix~A, Proposition~1). 

\subsection{``Prediction'' of the distorted percept, and shape of the distortion}

The exact geodesic $\gamma_\alpha$ is not known explicitly, and we have to rely on approximations to this first candidate for the prediction of the distorted percept. Here we outline the main steps leading to our final prediction, thereby leaving out the mathematical details in order to simplify the presentation. Those details are fully treated in the appendices B, C, and D.

For definiteness, let the vector field $v$ be fixed, and let $\tau$ denote the target half circle which is supposed to be of radius $r>0$, with center at the origin.

\medskip\noindent
{\em Shape of the distortion.}
For $\alpha=0$ the target half circle $\tau$ is an exact geodesic in the $H_0 \equiv H$ metric, namely $\tau = \gamma_0$. Since the visual distortion is expected to be small, though discernible, the effect magnitude $\alpha$, and likewise, the deflection of the geodesic $\gamma_\alpha$ from $\tau$, should be small, too. Therefore, it is plausible to try an {\em ansatz} 
\[
\wht \doteq \tau + \alpha \sigma =: \wtt
\]
for the prediction $\wht$ of the distorted percept, with $\sigma$ arising as the limit of $(\gamma_\alpha - \tau)/\alpha$ as $\alpha \to 0$. As such, $\sigma$ only depends on $v$ and $\tau$, not on $\alpha$; hence we call $\sigma$ the (approximate) {\em shape} of the deflection. It is determined by another differential equation similar to the one characterizing $\gamma_\alpha$ (Appendix~C, Proposition~2).

\medskip\noindent
{\em Adding the lower part.} 
Thus far, we only considered the part of the stimulus figure lying in the upper half plane and described the (preliminary) prediction $\wtt \equiv \wtt_+$ for the percept of the upper half of the target circle. This is sufficient if the figure is symmetrical about the $x_1$ axis. Otherwise we proceed as follows: the lower part is mirrored at the $x_1$ axis into the upper half plane and is processed there in the same way as the upper part; after modification, it is mirrored back to the lower half plane, yielding the prediction $\wtt_-$ for the percept of the lower part of the target circle. By construction, the two parts meet at their endpoints and together give the preliminary prediction $\wtt = \wtt_- \cup \wtt_+$ for the full target circle. 

\medskip\noindent
{\em Area conservation.} 
We can now state the {\em final form,} for given parameter $\alpha$, {\em of our ``prediction''} $\wht\equiv \wht_\alpha$ for the distorted percept of the full target circle:
\[
\wht = \kappa\, \wtt 
\]
where the scale factor $\kappa \equiv \kappa_\alpha > 0$ is chosen such that the area enclosed by the curve $\wht$ equals $r^2\pi$, the area of the target circle. The purpose of this renormalization is to adjust for potential size effects that have nothing to do with the perceptual distortion induced by the context. The quotation marks in ``prediction'' shall once more emphasize the circumstance that our approach yields a proper prediction only for the shape of the distortion; the parameter $\alpha$ measuring its strength yet has to be determined experimentally.

% ----------------------------------------------------------------------
\section{Experimental demonstration}
\label{exp}

The purpose of the reported pilot experiment was (a)~to verify qualitatively the prediction of the context-induced distortion shape following from our model, and (b)~to assess quantitatively the magnitude of the distortion effect as measured by the parameter $\alpha$. According to the regression to orthogonality hypothesis \RO, $\alpha$ should be nonnegative, so that a prevalence of positive estimates $\widehat \alpha$ can be seen as empirical support for \RO. One may expect that $\alpha$ actually depends on a number of factors such as context, participant, specific experimental conditions, etc. A study of such systematic variations is beyond the present scope, however, and we will focus on the sign of the $\widehat \alpha$s only. 

For the experimental determination we employed again the method of {\em compensatory measurement}, where the magnitude of the ``counter-distortion'' for which the perceptual distortion vanishes serves as a measure of the real distortion. The rationale is obvious: if $\tau$ is perceived as $\tau+\alpha\,\sigma$, then (for small $\alpha$) $\tau-\alpha\,\sigma$ will approximately be perceived as~$\tau$ \citep[cf.][p.~409]{EW2012}.\footnote{%
This method has a long tradition in the {\GOI} research: Z\"ollner was probably the first to use it for the measurement of the context-induced tilt of straight lines, another {\GOI} now named by him \citep{Zoe1860,Zoe1872}. For another use of this method (dubbed ``nulling paradigm'') in a dynamical, motion induced visual illusion cf.\ \citet{YaGo2011}.}
The procedure based thereupon is obvious as well: a series of circular targets distorted in the opposite direction is presented to the observer, who adjusts $\alpha$ until s/he perceives a perfect circle.

\begin{figure}[ht] 	%%%%%%%	Figure 3 %%%%%%%
\putfig{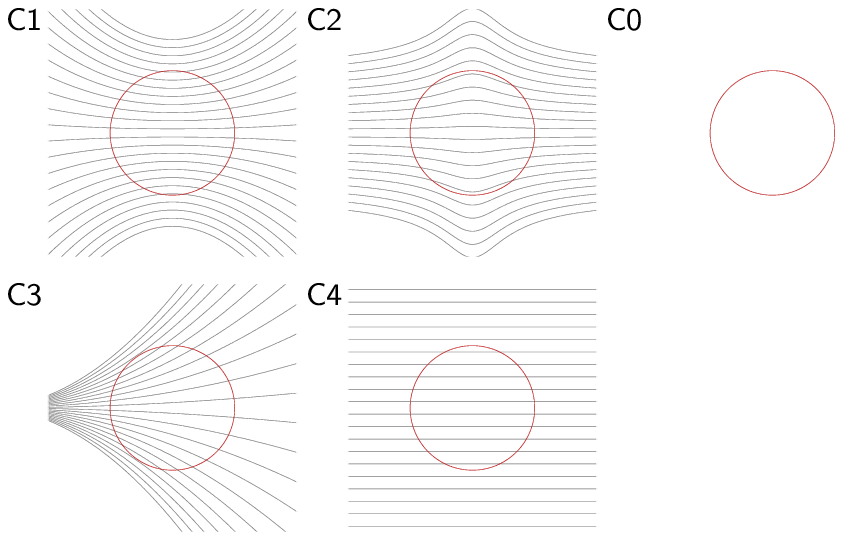}{1}
\hspace*{2mm} 
\parbox[b]{5.0cm}{
\small
Figure~3. Stimuli used in the experiment. Red closed curves are {\em exact} circles ($\alpha = 0$), superposed on context patterns (grey, C1--C4) or on a blank background (C0).
\vspace*{21ex}
}
\end{figure}

The stimuli were constructed for four different contexts C1 to C4, as shown in Fig.~3 and specified in Appendix~E. For each context, a series of 31 pictures was generated, showing the same array of context curves superposed with differently deformed targets of the form $\wht_{\alpha_k} = \tau - \alpha_k\,\sigma$, $(k=1,\dots,31),$ with coefficients $\alpha_k$ varied in a sufficiently large range. As a no-context control condition (C0), another stimulus series was constructed consisting of 31 pictures of ellipses on a blank background, with the main axes aligned with the horizontal and vertical axis, respectively, and half-axes ratios $r_2/r_1$ varied as $1 + \alpha_k$ while keeping the area constant across the picture series. This condition was included to check for a possibly confounding effect of the ``vertical--horizontal illusion'' (\VHI) as explained in Section 4.1. Details regarding the preparation of the stimuli and the experimental design are given in the supplementary online material.

Eight observers participated in the experiment. The contexts C1 to C4 were shown either in their original orientation (symmetry axis horizontal) or rotated by 90$^\circ$, with two repetitions each. Interspersed were five control trials using the no-context stimuli C0, summing up to 21 stimulus presentations in a single session. We thus obtained a total of 168 parameter estimates $\widehat\alpha$. The complete set of estimates $\widehat\alpha$ is presented in Fig.~4.\footnote{%
The results for the control C0 are displayed together with those for the contexts C1 to C4 only for parsimony of presentation; a direct, quantitative comparison of the estimates across the conditions is not meaningful.} 
Evidently, the average $\widehat\alpha$s for each pair of a subject and a (genuine) context are positive, with one single exception among the 32 cases (4 contexts $\times$ 8 subjects). This indicates that, statistically, the direction of the perceptual distortion is correctly predicted by our approach. That not all $\alpha$ estimates are positive, as it was the case in our earlier study of straight line targets \citep{EW2012}, is of little surprise given the small effect magnitude on the one hand, and the considerable intra-individual variance of the responses on the other hand. 

\begin{figure}[ht] 	%%%%%%%	Figure 4 %%%%%%% 
\begin{center}
\putfig{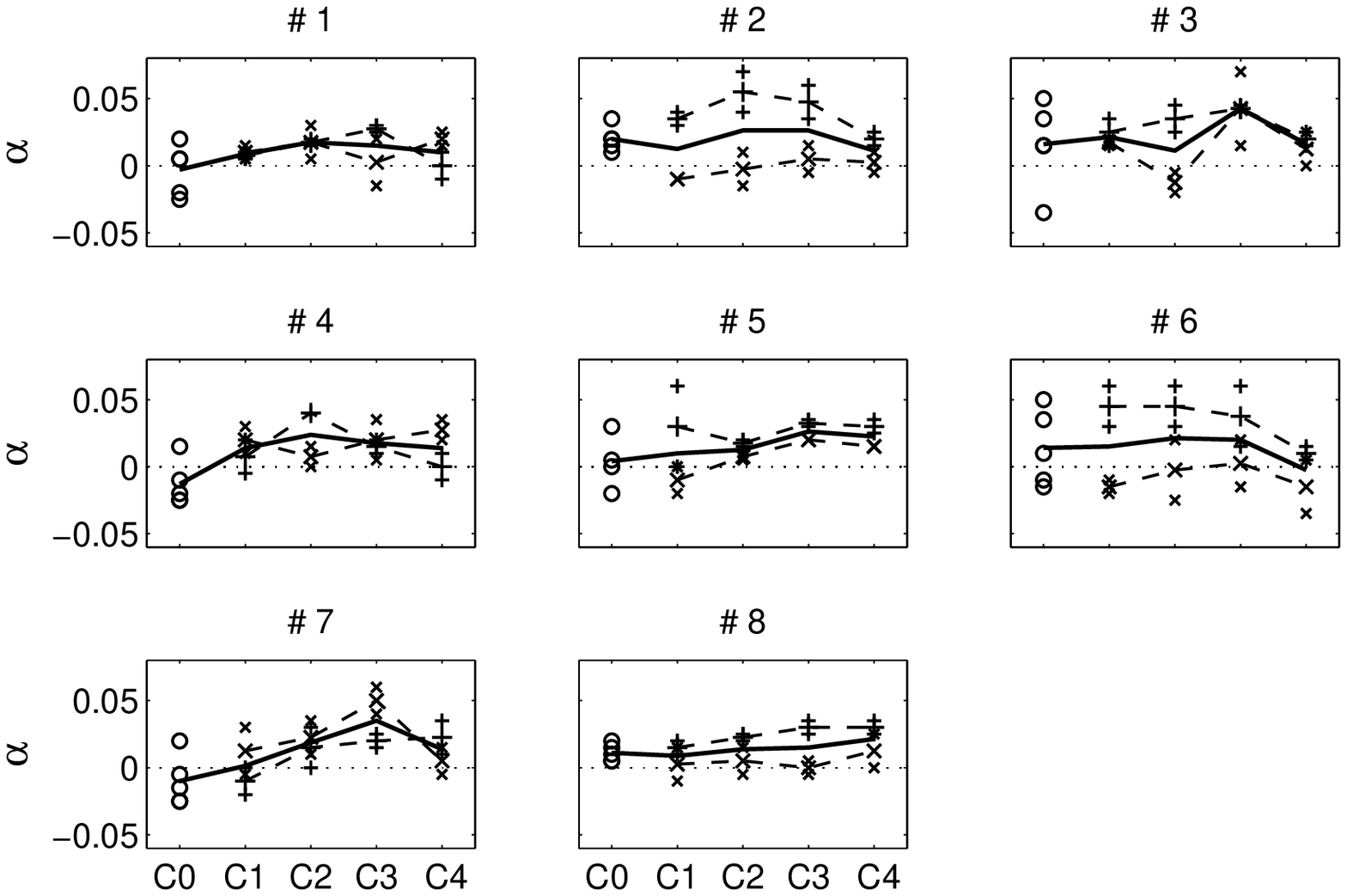}{0.833333}
\end{center}
\small 
Figure~4. Results of the experiment. For each subject, the single-trial $\alpha$ estimates are plotted vs.~control condition C0 and context patterns C1 to C4. Different markers are used for C0 (`$\circ$') and the contexts C1 to C4 (`$\times$'). The average of the single trial $\widehat\alpha$s is marked by a diamond for condition C0, and by a solid line for the others. 
\end{figure}

As for the possible influence of the {\VHI} (see Section \ref{ppc}), there is little evidence for a systematic overestimation of vertical extents: in the control condition C0, 5~out of~8 of the average $\widehat\alpha$s are positive (fraction of single trial positive $\widehat\alpha$ = 0.625, corresponding to $p \approx 0.08$, one-sided sign test). For the contexts C1 to C4 the tendency for positive $\alpha$ estimates is significantly higher (31~out of~32 of the average $\widehat\alpha$s are positive; fraction of single trial positive $\widehat\alpha$ = 0.766, corresponding to $p < 10^{-9}$, one-sided sign test). It is thus unlikely that the results of the experiment could be reduced to a vertical  vs.\ horizontal bias.%
\footnote{The case of the context C4, consisting of parallel horizontal lines, deserves an extra comment. As indicated by average $\widehat\alpha > 0$, the target subjectively appeared larger in the vertical than in the horizontal extension, which---as just argued---was not due to a vertical vs.\ horizontal bias. This finding seems to be in agreement with the ``filled space expansion,'' special cases of which are the Oppel--Kundt illusion or Helmholtz squares \citep{Rob1998,Wac2011b}. On the other hand, the perceived vertical elongation is in striking contrast with the familiar dressmaker's wisdom: to appear slim avoid wearing clothes with horizontal stripes \citep[cf.\ also][]{Tho2011}.}

% ----------------------------------------------------------------------
\section{Discussion}

The similarities and differences of our geometric modeling approach with the related work of Hoffman (\citeyear{Hof1966}, \citeyear{Hof1971}), Smith (\citeyear{Smi1978}), and Zhang and Wu (\citeyear{ZhWu1990}) were already expounded in our earlier communication \citep{EW2012}. Here we will focus on several other issues of interest that were not, or only marginally, addressed therein.

\subsection{Problem of ``phenomenal correctness''}
\label{ppc}

The modeling approach outlined in Section~2 yields, for a given context, a prediction for the shape of the distorted circular target. Obviously, we cannot exclude the existence of alternative shapes yielding a still better perceptual fit to a perfect circle. Naturally then, the question arises how to assess the merit of our prediction? It should be clear that a fully satisfactory answer is out of reach, on the following grounds.

In contrast with some other {\GOI}s in which only one parameter of the stimulus is affected (\eg\ length of a line segment, as in the M\"uller-Lyer illusion), the space of possible deformations of a circle into ovals and other closed curves is of infinite dimension. As it is thus impossible to exhaust all possible deformations, only a limited selection of (counter-)distortions could be presented to the observers for a comparative judgment. This requires, on the one hand, suitable theories or models delivering meaningful candidates for such a selection, that ought to incorporate the variation of the perceptual distortions with the context. Here our approach could be most helpful, as it specifies a concrete context-specific shape of the distortion. On the other hand, the sheer difficulty of having to distinguish between very similar oval-shaped forms sets substantial limitations to experimental approaches. As attested by the variability in our $\alpha$ estimates, such a task presents a serious challenge to the observers' perceptual faculties even when, as here, the shape is fixed and only one parameter (magnitude) is to be adjusted.

One may wonder whether, at least with sufficiently symmetrical contexts, a simple comparison between the figure's extent along its horizontal and vertical axes might also provide a convenient cue as to a deviation from circularity. It is known since the very beginnings of \GOI\ research \citep{Opp1855} that linear extents in the vertical direction are overestimated relatively to the horizontal direction \citep[p.~96--100]{Rob1998}. This so-called ``vertical--horizontal illusion'' (\VHI) is usually demonstrated in T-shaped or L-shaped line-drawn figures \citep{Leh1967,AD1969,Rob1998,Wol2005}, but it can also be observed in the perception of two-dimensional forms or three-dimensional real-world objects \citep{Met1975}. This was the rationale for including the control condition in our experiment, in order to evaluate the observers' ability to distinguish between circular and ellipsoidal forms, unaffected by a context. In fact, a {\VHI} effect could easily be incorporated in our model by replacing
\[
I = \mfour{1}{0}{0}{1} \rightarrow \mfour{1}{0}{0}{\beta^{2}} \qquad \mbox{for some~} \beta > 0
\]
in the definition of the metric tensors (cf.~(\ref{metric}), (\ref{pmetric})). However, our experimental data provided little if any evidence for a significant \VHI\ influence (Section~\ref{exp}). It is thus doubtful that the flexibility gained by an additional parameter, $\beta$, would compensate for the further increased difficulty of the perceptual task.

Notwithstanding the difficulties, we can say that our approach yields predictions capturing essential features of the observed distortions. The experimental data (prevalence of positive $\alpha$ estimates) and direct observation corroborate this claim (cf.\ Fig.~5).

\begin{figure}[ht] 	%%%%%%%	Figure 5 %%%%%%%
\begin{center}
\putfig{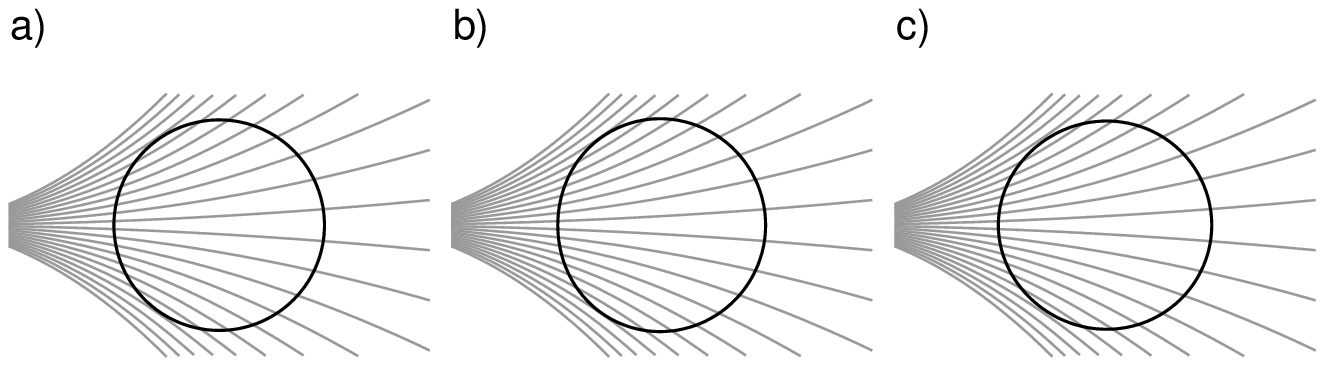}{1}
\end{center}
\small
Figure 5. Regarding the ``phenomenal correctness'' of the model-based prediction. Context C3 with predictions $\wht_\alpha = \tau + \alpha\,\sigma$ for $\alpha=0$ (a: perfect circle), $\alpha=+0.03$ (b: the indenting of the target at the left-hand side is strengthened \wrt\ the ``pure illusion'' in case a), and $\alpha = -0.03$ (c: counterbalancing yields an apparently circular shape).
\end{figure}

\subsection{Geometry of visual space}

Our geometric paradigm may be reminiscent of studies of the metric properties of visual space, a topic at times intensely debated in the vision research. The hypothesis of a non-Euclidean structure of visual space, introduced by \citet{Lun1948,Lun1950}, elicited a respectable amount of theoretical \citep{Sup1977,Fre1987} and experimental work \citep[for reviews see][]{Ind1991,Wag2006,Wes2008}. Those studies, however, were concerned with the geometry of the ``full'' three-dimensional [3D] visual space, as exemplified by Blumenfeld's (\citeyear{Blu1913}) experiments with the reproduction of visually assessed spatial distances. Also, in attempts of explaining some {\GOI}s on the basis of a non-Euclidean geometry of visual space, differences between judgments in the ``near'' and ``far'' spaces play a r\^ole \citep{Hee1983}. In fact, studies of ``the'' geometry of visual space remained remarkably inconclusive \citep{Wag2006}, or at least they did not support the original hypothesis of a visual space described by a Riemannian geometry of constant curvature: ``Instead, the geometry of visual space itself appears to be a function of stimulus conditions'' \citep[p.~493]{Wag1985}. Again, this assertion may be remotely reminiscent of our own approach; however, there are substantial differences.

Firstly, the stimuli in our experiment are definitely 2D, a possible 3D interpretation notwithstanding. Hence, the present work is dealing with the perceived content of the visual {\em field}, understood as a 2D manifold of primary visual experience \citep{Wac2011a}.\footnote{Of course, the visual field can be conceived of as a basis upon which the ``full'' visual space is constituted. This, however, is a very complex issue involving not only facts of vision but also integration of optic, haptic, and kinesthetic data; that is, a topic definitely beyond the scope of the present work.} Secondly, and most importantly, the geometry here is implicit, as it were, in the geometrical form of the {\em target} (\eg, Euclidean for straight lines, hyperbolic for circles). Aspects of vision and perception enter only via the {\em distortion} of that ``base'' geometry induced by {\em additional elements} (besides the target stimulus) in the visual field, namely, the context curves. Consequently, although the {\GOI} phenomena are undoubtedly intertwined with the issues of visual space perception \citep{Wes2008}, there is no direct connection between our approach and the research into ``the'' geometry of visual space in general.

\subsection{Explanatory status of the present theory}
\label{explstatus}

The approach taken in the present and the earlier \citep{EW2012} study is expressly ``phenomenological,'' that is, we are not aiming at explanations via the underlying neural or psychological mechanisms. This is a legitimate choice, adopted by other authors in the field as well \citep[\eg][]{Hof1966,Smi1978,Wag1985}. Notwithstanding this confinement, our approach does contain elements contributing to a better understanding of the {\GOI}s under study.

The visual field, as well as the respective retinal or cortical areas, are 2D varieties. Our two principal assumptions, ``local interactions'' and ``regression to ortho\-gonality'' (Section \ref{taco}), would correspond to inhibitory lateral interactions between adjacent, orientation-specific elements \citep{Bur1971,Car1973}. They thus furnish a clue for connections with the neurophysiology-based approach. However, unlike neural models proposed by other authors \citep[\eg][]{Wal1973,FeMa2004}, our approach models the {\GOI} phenomena where they occur, ``out there'' in the visual field, not in the retina or cortical areas. 

This is not to say that the present approach does nothing for explanations of the {\GOI}s. We only have to avoid the ``mechanistic fallacy,'' where the explanation of a phenomenon is identified with its reduction to the mechanisms acting at the most fundamental level, and other kinds of theories are dismissed as ``merely descriptive.''\footnote{%
A clarification may be appropriate at this point. It is true that in the natural sciences the knowledge of ``mechanisms''---\ie, of the elementary constituents of a system and of the rules governing their interactions---is the desirable and ultimate goal. This, however, does not imply that the whole of the scientific discipline is stated in terms of those elementary mechanisms. Descriptions of phenomena in purely functional terms abstracting from the elementary level lead to working theories as well. Disciplines such as thermodynamics, chemical kinetics, etc., operate without direct reference to the microscopic structure of matter or to elementary interactions between the individual molecules or atoms.}
For example, Snell's law of refraction describing the path of light propagation in heterogeneous media is definitely a phenomenological law. As such, it is well able to explain large varieties of optical phenomena. But even more importantly, Snell's law and other facts of dioptrics and catoptrics can be subsumed under (or derived from) a single superordinated extremal principle, the Fermat's principle of least time. Explanations from such principles are not less ``causal'' than mechanistic explanations, but they rely upon a different notion of causality, namely that of the formal rather than the efficient cause.

The relevance of extremal principles for human perception was first pointed out by Ernst Mach (\citeyear{Mach1866}) who raised the question, ``why a straight retinal image is seen [in the outward space] also as straight.'' In a more general note, he further speculated:
\begin{quote}\small
``As is well-known, mechanics leads everywhere to [solving] minimization or maximization tasks. However, the importance of such minimal principles exceeds by far the special assumptions of mechanics. They occur everywhere where a multitude [of elements] acts collectively, partly by facilitation, partly by inhibition. Also the eye, in seeing forms, is ruled by such principles.'' \citep[p.~19]{Mach1868}
\end{quote}

Our study thus can be seen as a variation on this general theme \citep[cf.\ also][]{HaEp1985}: to identify an overarching extremal principle from which special instances of ``seeing the straight as straight''---or more generally, ``seeing the shortest as shortest (in a given disturbed geometry)''---can be derived. We may speculate that this principle can be expressed in terms of ``minimal energy'' or ``minimal effort'': under static conditions as in the {\GOI}s studied here, the visual system would relax to a kind of equilibrium state requiring the least effort needed to resolve a ``perceptual tension'' between the target figure and the context pattern. If a related macroscopic variable characterizing a given percept can be defined, it should find a counter-part in a model of the neural substrate on the mesoscopic or microscopic level. Our focus on a phenomenological theory thus does not preclude a connection to ``mechanistic'' models of {\GOI}s: the present study may open a way for a research program integrating two different levels of modeling and theory.

To be sure, the question of how the visual system deals with motion or the uncertainties of 3D vision is highly relevant biologically, and experiments involving such elements help to better understand a number of visual illusions in terms of the links to and the working of the neural substrate; see \eg\ \cite{YaGo2011}, \cite{GoSt2014}. Conversely, such knowledge can lead to important insights, concerning \eg\ the perceived integrity of an object under rigid motion \citep{ZhWu1990}. 
Again, it cannot be denied that the existing explanatory theories of {\GOI}s, specifically, lack consensus even about the most basic issues. Common explanations usually refer {\em either} to hypothetical cognitive processes {\em or} to physiological mechanisms \citep[see \eg][]{CoGi1978}. This duality reflects the principal controversy between ``empiricist'' and ``nativist'' approches to perception \citep[cf.][]{Bor1942,Tur1994}, and continues up to the present \citep[see \eg][]{Rog2014,Man2015}. The neurophysiology approach searches the basis of {\GOI}s and related phenomena in the intrinsic properties of the visual system \citep{LiHu1987,Eag2001}. Cognitive theories, by contrast, attempted to explain the {\GOI}s via ``unconscious inferences'' \citep{Hel1867,Rock1977,Gre1997a} based on additional hypotheses such as 3D interpretations of pictorial stimuli, as for example in Gregory's (\citeyear{Gre1963}) theory of ``inappropriate constancy scaling''. As for the 3D interpretation of {\GOI} stimuli: there is no evidence that a planar drawing is {\em necessarily} seen as a 3D scene \citep{Wac2011a}. The same can be said about other hypothetical constituents of those theories, such as ``imagined motions'' \citep{Ehr1925,Cha2008}. No doubt, there {\em are} phenomena involving 3D vision as well as illusory or real motion \citep[see \eg][]{YaGo2011}. However, the specific character of the {\GOI}s under study, which occur in the 2D visual field and are purely static in nature, furnishes no evidence, at this point, how those elements could be incorporated in our approach.

\subsection{Conclusion and outlook}

The present work extends our earlier work about geometric--optical illusions \citep{EW2012} from straight line targets to circle targets. Thereby we had to overcome some technical difficulties: since the geodesics in the appropriate (hyperbolic) geometry are {\em half} circles, we had to divide the stimulus in two parts and to recombine the distorted percepts. Another, rather technical issue concerned the problem of aligning the distorted percept to the target (Appendix~B). A certain degree of arbitrariness is always involved. However, the influence of those arbitrary choices appears to be immaterial. E.\,g., rotating the stimulus by $90^\circ$ has but a minor effect on the predicted percept while the major qualitative features of the distortion are maintained. Also, bisecting the stimulus along the $x_1$ axis is not unnatural as the horizontal axis {\em has} an outstanding direction, considering the position of the eyes (which were fixated during the experiment). Finally, in accordance with the ``local interactions'' hypothesis the illusory distortion is observable not only in full circles, but also when using circular {\em arcs} as targets.

The initial motive for the present work was to explore the scope of the geometrical approach adopted in our earlier paper on straight line targets. We found that the case of a circle target could be reconciled with the former framework by conceiving of the target itself as a geodesic in a suitable base geometry. The apparent commonalities of the two special cases lead us to propose the following general approach to the modeling of geometric--optical illusions of the ``target--context'' type (cf.\ Fig.~6):

{\em Consider (only) targets that are geodesics in some base Riemannian geometry associated with metric tensor $M$; model the illusory percept of the target as a geodesic in another geometry associated with a (small) context-dependent perturbation $\widetilde M$ of $M$} (which involves the ``local interactions'' and ``regression to right angles'' principles).

\begin{figure}[ht]	%%%%%%%	Figure 6 %%%%%%%
\begin{center}
\putfig{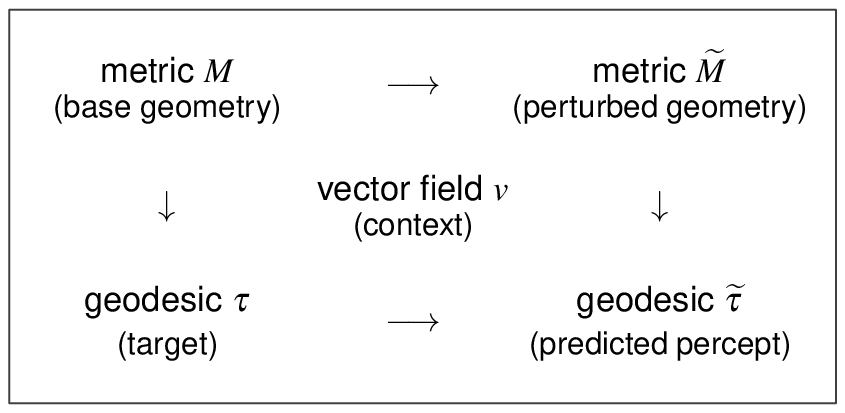}{1.25}
\end{center}
\small
Figure~6. Illustration of the general modeling principle in the form of a commutative diagram.
\end{figure}

Applications of this framework to other, more general types of targets may be expected to encounter specific problems. For example, in our experiments we were relying upon the observers' perceptual experience of the ``ideal form,'' enabling them to detect perceptual distortions. While this could be taken for granted for straight lines or circular forms, it may not apply to other forms, such as conics. The problem could be solved by presenting the ``variable'' target, superposed with the context pattern, along with a ``standard'' (non-distorted) form shown side-by-side, and instructing the subject to adjust the variable target to ``perceptual equality.'' Other challenges requiring further mathematical and computational work may occur in determining a suitable base geometry for a given target form. Nonetheless, we believe that the approach proposed in the present paper is a viable one and thus deserves further investigation and testing.

\section*{Acknowledgments}

Part of this work was done while the authors were with the Institute for Frontier Areas of Psychology and Mental Health, Freiburg i.~Br., Germany.  W.E.\, thanks the Klaus Tschira Foundation for support. We are grateful to Steffen Heinze for his important hint regarding the connection with Riemannian geometry, and Jakob Pacer for assistance in the experiments.  Finally, we thank the editor and three anonymous referees for their valuable comments and recommendations.

% ---------------------------------------------------------------------
\section*{Appendix}

\subsection*{A. Euler--Lagrange equation for the $\Ha$ geodesics }

The differential equations characterizing geodesic curves commonly are stated using Christoffel symbols derived from the metric tensor \citep{Lau1965}. A more intelligible, index-free form can be obtained from the Euler--Lagrange equations for the variational problem of minimizing the curve length $L_\alpha(x)$ within the set $\X$ of all twice continuously differentiable curves $t \mapsto x(t),\, t_0\leq t \leq t_1$ in $\hp$ with fixed, given endpoints $x(t_0) = \tau_0,\, x(t_1) = \tau_1$. For convenience we shall suppress the parameter $t$ as well as the parameter $\alpha$ wherever possible. Vectors in $\real^2$ are conceived (and treated) as column vectors, but written as row vectors when reference is made to the components. For instance, $\Fe_2 = (0,1)$ is the second basis vector in the standard basis of $\real^2$. The expression $v'(\xi)$ denotes the $2 \times 2$ matrix of first-order partial derivatives $\pd v_i(\xi)/\pd \xi_j$ of the vector field $v \equiv v(\xi)$ (Jacobian), and $\omega(\xi) = \pd v_1(\xi)/\pd \xi_2 - \pd v_2(\xi)/\pd \xi_1$ denotes its rotation (evaluated at $\xi$ each). 
The details of the derivation involve the following steps similar to \citep{EW2012} (henceforth: EW2012).

\medskip\noindent
{\sc Step~1}: {\em Equivalent minimization problem.} We first consider the more simple problem of minimizing the ``energy functional'' (cf.~Section \ref{explstatus})
\[
J(x) = \int_{t_0}^{t_1} F(x,\dx)\, dt 
\]
where 
\[
F(x,\dx) = \langle \dx, \Ha(x)\, \dx \rangle = x_2^{-2} \left[ |\dx|^2 + 2\alpha \langle \dx, v(x) \rangle^2 \right].
\]
A minimizing curve $x$ necessarily satisfies the (vectorial) {\em Euler--Lagrange equation}
\beq \label{eeq}
\frac{d}{dt}\, \nabla_\dx F(x,\dx) - \nabla_x F(x,\dx) = 0 ,
\eeq
wherein $\nabla_x F,\, \nabla_\dx F$ denote the partial gradients of $F$ with respect to the (vector) arguments $x, \, \dx$, respectively. For the present $F$ we have
\beqas
\nabla_\dx F \ceq x_2^{-2} \left[ 2 \dx + 4 \alpha \langle \dot x, v(x) \rangle\,v(x) \right],\\
\frac{d}{dt}\, \nabla_\dx F \ceq - 4\, \frac{\dx_2}{x_2^3} \left[\dx + 2 \alpha \langle \dot x, v(x) \rangle\,v(x) \right] \\
&& \quad +\, x_2^{-2} \left[2 \ddx + 4\alpha v(x) \, \frac{d}{dt} \langle \dx, v(x)\rangle + 4\alpha \langle \dx, v(x)\rangle \, v'(x) \dx \right], \\ 
\nabla_x F \ceq -\frac{2}{x_2^3}  \left[|\dx|^2 + 2\alpha \langle \dx, v(x) \rangle^2 \right] + 4\alpha\, \frac{\langle \dx, v(x)\rangle}{x_2^2}\, v'(x)^\ast \dx,
\eeqas
with $v'(x)^\ast$ the adjoint of $v'(x)$. 
With these expressions, the Euler--Lagrange equation becomes
\beqa
0 \ceq \frac{d}{dt}\, \nabla_\dx F(x,\dx) - \nabla_x F(x,\dx) \nonumber\\ \ceq
\frac{2}{x_2^2} \left[ \ddx + 2\alpha v(x) \, \frac{d}{dt} \langle \dx, v(x)\rangle + 2\alpha \langle \dx, v(x)\rangle \, v'(x) \dx - 2\alpha\, \langle \dx, v(x)\rangle\, v'(x)^\ast \dx \right] \nonumber\\ && \quad -\frac{2}{x_2^3} \left[ 2\dx_2 \left(\dx + 2\alpha \langle \dx, v(x)\rangle\, v(x) \right) - \left(|\dx|^2 + 2\alpha \langle \dx, v(x) \rangle^2 \right) \, \Fe_2 \right] \nonumber\\ \ceq
\frac{2}{x_2^2} \left[ \ddx + 2\alpha v(x) \, \frac{d}{dt} \langle \dx, v(x)\rangle + 2\alpha \langle \dx, v(x)\rangle  \left(v'(x) - v'(x)^\ast\right) \dx \right] \nonumber\\ && \quad -\frac{2}{x_2^3} \left[ 2\dx_2\, G_\alpha(x) \dx - \langle \dx, G_\alpha(x) \dx \rangle \, \Fe_2 \right] \label{ee0}
\eeqa
where $G_\alpha(\xi) = I + 2\alpha v(\xi) \otimes v(\xi)$ as in (\ref{pmetric}). 

We now show that if $t \mapsto x(t)$ is a solution of this equation, then $F(x(t),\dx(t))$ is constant in $t$. This follows by writing the Euler--Lagrange equation (\ref{eeq}) in DuBois--Reymond's form (\eg, \citep[Sect.~2.3]{Dac2004})
\[
\frac{d}{dt} \left(F - \langle \nabla_\dx F, \dx \rangle) \right) = \pd_t F
\]
and noting that $\pd_t F \equiv 0$ since $F$ does not depend explicitly on $t$. It follows that
\[
\langle \nabla_\dx F, \dx \rangle - F = x_2^{-2} \left[2 |\dx|^2 + 4\alpha \langle \dx, v(x) \rangle^2 \right] - F = 2F - F = F
\]
is indeed constant as a function of $t$. As an easy consequence, a stationary point of the functional $x \mapsto \int_{t_0}^{t_1} F(x,\dx)\, dt = J(x)$ is also a stationary point of the functional $x \mapsto \int_{t_0}^{t_1} \sqrt{F(x,\dx)}\, dt = L_\alpha(x)$, so in the following we may continue with the former.

\medskip\noindent
{\sc Step 2.} {\em Euler--Lagrange equation}. In order to state the Euler--Lagrange equation (\ref{ee0}) in explicit form, we use the fact taken from EW2012 that
\[
\left(v'(x) - v'(x)^\ast\right) \dx = \omega |\dx| \rho^\bot
\]
where $\rho = \dx/|\dx|$ is the direction of the curve $x$, $\rho^\bot$ is the unit vector orthogonal to $\rho$ making $\rho,\, \rho^\bot$ a positively oriented basis of $\real^2$, and $\omega = \pd_1 v_2 - \pd_2 v_1$ is the rotation of the vector field $v$ (evaluated at $t$ and $x(t)$, respectively). Multiplying (\ref{ee0}) with $x_2^2/2$ and suppressing the argument $x \equiv x(t)$ of $v$ and $G_\alpha$ gives
\beqas
0 \ceq \ddx + 2\alpha\, v\, \frac{d}{dt} \langle \dx, v\rangle + 2\alpha\, \langle \dx, v\rangle\, \omega |\dx| \rho^\bot - \frac{1}{x_2} \left[ 2\dx_2\, G_\alpha \dx - \langle \dx, G_\alpha \dx \rangle \, \Fe_2\right] \\ \ceq
\ddx + 2\alpha\, \langle \ddx, v\rangle\, v + 2\alpha\, \langle \dx, v' \dx \rangle\, v + 2\alpha\, \langle \dx, v\rangle\, \omega |\dx| \rho^\bot - \frac{1}{x_2} \left[ 2\dx_2\, G_\alpha \dx - \langle \dx, G_\alpha \dx \rangle \, \Fe_2\right] \\ \ceq 
G_\alpha \ddx  + 2\alpha\, \langle \dx, v' \dx \rangle\, v + 2\alpha\, \langle \dx, v\rangle\, \omega |\dx| \rho^\bot - \frac{1}{x_2} \left[ 2\dx_2\, G_\alpha \dx - \langle \dx, G_\alpha \dx \rangle \, \Fe_2\right].
\eeqas
By inversion using $G_\alpha^{-1} = I - \frac{2\alpha}{1+2\alpha}\, v \otimes v$ we obtain
\beqa \label{ee1}
\ddx \ceq \frac{2 \dx_2}{x_2}\, \dx - \frac{|\dx|^2 + 2\alpha\, \langle \dx, v\rangle^2}{x_2} \left[ \Fe_2 - \frac{2\alpha  v_2}{1+2\alpha}\, v\right] \\ && \qquad\ 
- 2\alpha |\dx|^2 \left[ \frac{\langle \rho, v' \rho \rangle}{1+2\alpha}\, v + \omega \langle \rho, v\rangle \left( \rho^\bot - \frac{2\alpha}{1+2\alpha}\, \langle \rho^\bot, v\rangle\, v \right) \right], \nonumber
\eeqa
which is the Euler--Lagrange equation in explicit form.

\medskip\noindent
{\sc Step 3}: {\em Reparameterization}. To proceed we switch to the alternative parameterization 
\[
s = \tanh t, \qquad z(s) = x(\mathrm{arctanh}\, s).
\]
We have $\frac{ds}{dt} = 1-\tanh^2 t \equiv 1-s^2$, hence 
\[
x(t) \equiv z(s),\quad \dx(t) = \frac{dz}{ds}\, \frac{ds}{dt} \equiv z'(s) (1-s^2),\quad \ddx(t) \equiv z''(s) (1-s^2)^2 - 2z'(s) s(1-s^2).
\]
Moreover,
\[
\frac{\dx(t)}{|\dx(t)|} \equiv \frac{z'(s)}{|z'(s)|} \equiv \rho,\qquad |\dx(t)|^2 \equiv |z'(s)|^2 (1-s^2)^2.
\]
With these substitutions the Euler--Lagrange equation (\ref{ee1}) becomes
\beqas
&& \!\!\! (1-s^2)^2\, z'' - 2s (1-s^2)\, z' \\ \ceq 2\, (1-s^2)^2\, \frac{z_2'}{z_2}\, z'\, -\, (1-s^2)^2\ \frac{|z'|^2 + 2\alpha \langle z',v\rangle^2}{z_2} \left[ \Fe_2 - \frac{2\alpha  v_2}{1+2\alpha}\, v\right] \\ && 
-2\alpha (1-s^2)^2 |z'|^2 \left[ \frac{\langle \rho, v' \rho \rangle}{1+2\alpha}\, v + \omega \langle \rho, v\rangle \left( \rho^\bot - \frac{2\alpha}{1+2\alpha}\, \langle \rho^\bot, v\rangle\, v \right) \right],
\eeqas
or after division by $(1-s^2)^2$,
\beqas
z'' \ceq 2 \left( \frac{z_2'}{z_2} + \frac{s}{1-s^2} \right) z'\, -\, \left(1 + 2\alpha \langle \rho,v\rangle^2 \right) \frac{|z'|^2}{z_2}\, \left[ \Fe_2 - \frac{2\alpha  v_2}{1+2\alpha}\, v\right] \\ && 
-2\alpha\, |z'|^2 \left[ \frac{\langle \rho, v' \rho \rangle}{1+2\alpha}\, v + \omega \langle \rho, v\rangle \left( \rho^\bot - \frac{2\alpha}{1+2\alpha}\, \langle \rho^\bot, v\rangle\, v \right) \right].
\eeqas
Regrouping terms according to the vectors involved we obtain the Euler--Lagrange equation in the form used here, stated as Eq.~(\ref{elz}) in the following.

\medskip\noindent
{\bf Proposition 1. } {\em 
Consider the variational problem of minimizing the curve length $L_\alpha(z)$ within the set of all twice continuously differentiable curves $s \mapsto z(s),\, s_0\le s \le s_1$ in $\hp$ with fixed, given endpoints $z(s_0) = \tau_0,\, z(s_1) = \tau_1$. If $z$ is a solution to this problem, it satisfies the (vectorial) Euler--Lagrange equation}
\beqa \label{elz}
z'' \ceq 2 \left( \frac{z_2'}{z_2} + \frac{s}{1-s^2} \right) z'\, -\left(1 + 2\alpha \langle \rho,v\rangle^2 \right) \frac{|z'|^2}{z_2}\ \Fe_2\, -\, 2\alpha |z'|^2\, \omega \langle \rho, v\rangle\, \rho^\bot \\ && 
-\,\frac{2\alpha}{1+2\alpha}\, |z'|^2 \left[ \langle \rho, v' \rho \rangle 
- \left(1 + 2\alpha \langle \rho,v\rangle^2 \right) \frac{v_2}{z_2} -2\alpha \omega \langle \rho, v\rangle \langle \rho^\bot, v\rangle \right]\! v. \nonumber
\eeqa

Let us remind here of the tacit understanding that the terms $z, z', z'', \rho, \rho^\bot$ are to be evaluated at the (suppressed) argument $s$, and $v, v', \omega$ at $z(s)$, respectively. The ambiguity in the meaning of the primes in $z'$ and $v'$, which denote differentation \wrt\ the arguments $s$ and $\xi$, respectively, will not cause confusion.

It is straightforward to check that for $\alpha=0$, the vertical lines in $\hp$ and the half circles orthogonal to the $x_1$ axis are indeed solutions to (\ref{elz}). In particular, any half circle of the form 
\beq\label{tau}
\tau(s) = r \left(s, \sqrt{1-s^2}\right), \qquad |s| < 1,
\eeq
playing the role of a target stimulus, is a geodesic in the $H_0$ geometry.

\subsection*{B. Numerical solution}

As in EW2012, we use a Picard--Lindel\"of type iteration scheme with endpoints fixed at those of the target. Naturally, one would like to have a complete half circle as the target. There are two problems, however. First, the length of a complete half circle is infinite in the given geometry; second, its endpoints are singular points of the differential equation. 

Our workaround is as follows. For given radius $r>0$, pick $s^* \in (0,1)$ close to 1, and take the points $\tau_0^* = \tau(-s^*),\, \tau_1^* = \tau(s^*)$ on the target (\ref{tau}) as the fixed endpoints for the iterative scheme. In that scheme, a sequence of planar curves $s \mapsto z_n(s)$, approximate solutions to (\ref{elz}) with endpoints at $\tau_0^*,\, \tau_1^*$, is defined as follows: for $|s| \le s^*$ and $n\ge 1$
\beqa\label{pic0}
z_0(s) \ceq \tau(s) \\
z_n'(s) \ceq \int_0^s Q(u,z_{n-1}(u),z_{n-1}'(u)) \, du + a_n \label{picnp}\\
z_n(s) \ceq \int_0^s z_n'(u) \, du + b_n \label{picn}
\eeqa
where $Q(s,z,z')$ is given by the right-hand side of (\ref{elz}) (with $s,z,z'$ substituted by, respectively, $u,\, z_{n-1}(u),\ z_{n-1}'(u)$), and the constants $a_n,\, b_n \in \real^2$ are chosen so that $z_n$ satisfies the boundary conditions:
\beqa\label{aconst}
a_n \ceq \frac{\tau_1^* - \tau_0^* - (\Delta_n^+ - \Delta_n^-)}{2s^*}\\
b_n \ceq \frac{\tau_1^* + \tau_0^* - (\Delta_n^+ + \Delta_n^-)}{2} \label{bconst}
\eeqa
where
\[
\Delta_n^\pm = \int_0^{\pm s^*} \int_0^v Q(u,z_{n-1}(u),z_{n-1}'(u)) \, du\, dv.
\]
Finally, the curves $z_n$ are continued beyond the range $|s| \le s^*$ by setting
\[
z_n(s) = \tau(s) \quad \mbox{for} \quad s^* \le |s| \le 1.
\]

Extensive checks showed that this procedure yields numerically robust results eventually independent of the choice of $s^*$ provided $s^*$ is chosen sufficiently close to 1; we ultimately took $s^* = 1-10^{-6}$. Note that if the limit $z_\infty$ of the above scheme exists, it will satisfy Eq.~(\ref{elz}), hence represent (for $|s| \le s^*$) a geodesic connecting the points $\tau_0^*,\, \tau_1^*$. The convergence properties of the scheme are briefly discussed in the next appendix. A rigorous convergence proof for an analogous iterative procedure was given in EW2012 for the case of straight line targets.

\subsection*{C. Approximate shape of the perceptual distortion}

As anounced in Section 2.3, the exact geodesic $\gamma_\alpha$ in the perturbed metric $\Ha$ will be approximated for small $\alpha$ by a curve of the form $\tau + \alpha \sigma$, where $\sigma$ depends on the vector field $v$ along $\tau$ only. In particular, $\sigma$ is independent of parameter $\alpha$, and it represents the {\em shape} of the perceptual distortion. We derive a differential equation characterizing $\sigma$ by letting the parameter $\alpha$ tend to zero in (\ref{elz}). 

In the following we ignore terms of the order $O(\alpha^2)$. This is indicated by replacing the equality sign by `$\doteq$'. Moreover, the quantities $v, v', \rho, \rho^\bot, \omega$ are understood to be evaluated at $z\equiv\tau(s)$. Making the {\em ansatz} $z \doteq \tau + \alpha \sigma$ we get for the single components
\[
z_1 \doteq rs + \alpha \sigma_1,\quad z_1'\doteq r + \alpha \sigma_1', \quad z_1''\doteq \alpha \sigma_1'',
\]
\[
z_2 \doteq r\sqrt{1-s^2} + \alpha \sigma_2,\quad z_2'\doteq -\frac{rs}{\sqrt{1-s^2}} + \alpha \sigma_2', \quad z_2''\doteq -\frac{r}{(1-s^2)^{3/2}} + \alpha \sigma_2''.
\]
Therefore,
\beqas
\frac{z_2'}{z_2} & \!\! \doteq \!\! & \frac{-\frac{rs}{\sqrt{1-s^2}} + \alpha\, \sigma_2'} {r\sqrt{1-s^2} \left(1 + \frac{\alpha\, \sigma_2}{r\sqrt{1-s^2}} \right)} \doteq -\frac{s}{1-s^2} + \frac{\alpha\,\sigma_2'}{r\sqrt{1-s^2}} + \frac{\alpha\, s\,\sigma_2}{r(1-s^2)^{3/2}} \\
& \!\! = \!\! & -\frac{s}{1-s^2} + \frac{\alpha}{r\sqrt{1-s^2}} \left(\sigma_2' + \frac{s}{1-s^2}\,  \sigma_2\right),
\eeqas
and similarly
\beqas
|z'|^2 & \!\! \doteq \!\! &  \frac{r^2}{1-s^2} + 2\alpha\, r \left( \sigma_1' - \frac{s}{\sqrt{1-s^2}}\, \sigma_2' \right),\\
\frac{|z'|^2}{z_2} & \!\! \doteq \!\! & \frac{r}{(1-s^2)^{3/2}} + 2\alpha  \left(\frac{\sigma_1'}{\sqrt{1-s^2}} - \frac{s}{1-s^2}\, \sigma_2' - \frac{\sigma_2}{2(1-s^2)^2}\right).
\eeqas
Inserting these expressions into the Euler--Lagrange equation (\ref{elz}) we obtain for the first and second components
\[
\alpha \sigma_1'' \doteq \frac{2\alpha}{r\sqrt{1-s^2}} \left[\sigma_2' + \frac{s}{1-s^2}\, \sigma_2' \right] (r + \alpha \sigma_1') - \frac{2\alpha\, r^2}{1-s^2} \left[ \left( \langle \rho,v' \rho\rangle - \frac{v_2}{\tau_2}\right) v_1 + \omega \langle \rho, v\rangle\, \rho_1^\bot \right]
\]
and
\beqas
-\frac{r}{(1-s^2)^{3/2}}\, + \alpha \sigma_2'' & \!\! \doteq \!\! & \frac{2\alpha}{r\sqrt{1-s^2}} \left[\sigma_2' + \frac{s}{1-s^2}\, \sigma_2' \right] \frac{-rs}{\sqrt{1-s^2}} - \frac{r}{(1-s^2)^{3/2}}\\ && \!\!
- 2\alpha\, \left(\frac{\sigma_1'}{\sqrt{1-s^2}} - \frac{s}{1-s^2}\, \sigma_2' - \frac{\sigma_2}{2(1-s^2)^2}\right)  - \frac{2\alpha\, r\, \langle\rho, v\rangle^2}{(1-s^2)^{3/2}}\\ && \!\!
- \frac{2\alpha\, r^2}{1-s^2} \left[ \left( \langle \rho,v' \rho\rangle - \frac{v_2}{\tau_2}\right) v_2 + \omega \langle \rho, v\rangle\, \rho_2^\bot \right],
\eeqas
respectively. Regrouping terms, dividing by $\alpha$, then ignoring terms of order $O(\alpha)$ gives the following. 

\medskip\noindent
{\bf Proposition 2. } {\em The shape $\sigma = (\sigma_1,\sigma_2)$ of the perceptual distortion satisfies the following system of differential equations,}
\beqa \label{siga1}
\sigma_1'' \ceq \frac{2}{1-s^2} \left[ \sqrt{1-s^2}\, \sigma_2' + \frac{s}{\sqrt{1-s^2}}\, \sigma_2 - r^2 v_1 \left(\langle \rho,v' \rho\rangle - \frac{v_2}{\tau_2} \right) - r^2 \omega \langle \rho, v\rangle\, \rho_1^\bot \right], \\
\sigma_2'' \ceq - \frac{2}{1-s^2} \left[ \sqrt{1-s^2}\, \sigma_1' + \frac{s^2-1/2}{1-s^2}\, \sigma_2 + \frac{r\langle \rho, v \rangle^2}{\sqrt{1-s^2}} \right. \label{siga2}\\ && \qquad\qquad \left. +\, r^2 v_2 \left(\langle \rho,v' \rho\rangle - \frac{v_2}{\tau_2} \right) + r^2 \omega \langle \rho, v\rangle\, \rho_2^\bot \right]. \nonumber
\eeqa

We remind here of our stipulation that $v$, e.g., is evaluated at $z(s) = \tau(s)$, and likewise for the other quantities, $v', \rho,\rho^\bot, \omega$. Thus all terms besides $\sigma$ and its derivatives are known once the vector field $v$ and the target $\tau$ (i.e., the radius $r$) are fixed, meaning that the shape can indeed be derived from the system (\ref{siga1}), (\ref{siga2}). In the case of a straight line target treated in EW2012, this calculation reduced to just (twofold) integration. Here we have to solve a proper differential equation (system) which is, again, accomplished by Picard--Lindel\"of iteration as described in the previous appendix, setting $\sigma$ and $\sigma'$ identically 0 in the beginning. 

These algorithms for solving the equations in Propositions 1 and 2 converged for all contexts considered in our study, uniformly in the range $|\alpha| \le 0.1$ (which is sufficiently large since all $\alpha$ estimates in the experiment were $\le .07$). Convergence was not as rapid as for straight line targets, yet stationarity virtually was reached after 11 iterations. We therefore took the $N=11$-th iterates $z_N \equiv z_{\alpha,N}$ and $\sigma_N$ of the respective Picard--Lindel\"of schemes as substitutes for the exact geodesic $\gamma_\alpha$ and shape $\sigma$. Finally, discrepancies between the curves $z_{\alpha,N}$ and $\widetilde z_{\alpha,N} = \tau + \alpha \sigma_N$ were minimal to indiscernible by eye in all cases. 

\subsection*{D. Special case of a context consisting of horizontal lines}

Explicit solutions to (\ref{siga1}), (\ref{siga2}) generally are not available. An interesting exception is the case of a context consisting of horizontal lines. Then $v_1 \equiv 1,\, v_2 \equiv 0$, say, and $v' \equiv 0,\, \omega \equiv 0$. Since the components of $\rho$ evaluated along $\tau$ are $\rho_1 = \sqrt{1-s^2},\, \rho_2 = -s$, the system (\ref{siga1}), (\ref{siga2}) reduces to 
\beqa \label{siga1cv}
\sigma_1'' \ceq \frac{2}{1-s^2} \left[ \sqrt{1-s^2}\, \sigma_2' + \frac{s}{\sqrt{1-s^2}}\, \sigma_2 \right], \\
\sigma_2'' \ceq - \frac{2}{1-s^2} \left[ \sqrt{1-s^2}\, \sigma_1' + \frac{s^2-1/2}{1-s^2}\, \sigma_2 + r\sqrt{1-s^2} \right]. \label{siga2cv}
\eeqa

To study this case it is convenient to assume that the shape $\sigma$ is of the form $\sigma = \tau\,\ts$ (coordinatewise multiplication), so that in view of $z \doteq \tau (1+\alpha\,\ts)$ the curve $\ts$ represents a kind of multiplicative shape. With this substitution the system (\ref{siga1cv}), (\ref{siga2cv}) becomes 
\beqa \label{sigm1cv}
\ts_1'' \ceq \frac{2}{s} \left(\ts_2' - \ts_1' \right), \\
\ts_2'' \ceq \frac{2}{1-s^2} \left(\ts_2 - \ts_1 - 1 + s (\ts_2' - \ts_1') \right). \label{sigm2cv}
\eeqa

Now, subtracting (\ref{sigm1cv}) from (\ref{sigm2cv}) and putting $\eta = \ts_2 - \ts_1 - 1$ we get the single differential equation 
\[
\eta'' = \frac{2}{s(1-s^2)} \left(s \eta + (2s^2-1) \eta'\right),
\]
the general solution of which is $\eta = A \lambda + B/s$ where
\beq\label{lambda}
\lambda(s) = \frac{1}{s}\, \log \frac{1+s}{1-s},
\eeq
and $A, B$ are constants. Clearly $B=0$, since otherwise one would have a pole at $s=0$. Direct integration of (\ref{sigm1cv}) gives at first
\[
\ts_1'(s) = \int_0^s \frac{2\eta'(u)}{u}\, du + C = A\, \frac{(1+s^2)\, \lambda(s) - 2}{s} + C,
\]
then
\[
\ts_1 = A \left(2 - (1-s^2)\, \lambda\right) + Cs + D,
\]
and thus 
\[
\ts_2 = \eta + \ts_1 + 1 =  A \left(s^2\, \lambda + 2\right) + Cs + D + 1.
\]
The constants $C, D$ can be determined from the condition that $\sigma_1(s) = r s \ts_1(s)$, and hence $\ts_1(s)$, has to vanish at $s= \pm 1$. Thus $2A -C + D = 0,\, 2A + C + D =0$, and so $C=0,\, D = -2A$. Perhaps surprisingly, there appears to be no condition fixing the value of $A$. Note that the condition $\sigma_2(s) = r\sqrt{1-s^2}\, \ts_2(s)=0$ at $s = \pm 1$ does not help since the factor $\sqrt{1-s^2}$ vanishes there.

Summarizing, we have determined the shape $\sigma$ of the perceptual distortion in the case of a horizontal lines context up to a constant $A>0$: its components are 
\beq\label{sigcv}
\sigma_1(s) = rs A (s^2-1) \lambda(s), \quad \sigma_2(s) = r\sqrt{1-s^2}\left(1 + A s^2 \lambda(s) \right),
\eeq
where $\lambda$ is defined by (\ref{lambda}).  It should be noted that the resulting prediction $z^*_{\alpha,A} = \tau + \alpha \sigma$ (with $\sigma \equiv \sigma_A$ from (\ref{sigcv})) for the distorted percept is not just an ellipse. Remarkably also, the curves $z^*_{\alpha,A}$ are for $A = 1, 1/8, 1/32$ virtually indistinguishable from each other, and also from the approximations $z_{\alpha,N}$ and $\widetilde z_{\alpha,N}$ considered in Appendix C. 

\subsection*{E. Contexts used in the experiment}

A context consists of a finite sample from a family of curves $u \mapsto C_\theta(u)$ depending on a real parameter $\theta$. We specifically consider curves of the form $C_\theta(u) = (u,\theta q(u))$, where $q$ is a smooth function positive on a domain enclosing the domain covered by the target. The components of the associated (normalized) vector field $v = (v_1,v_2)$ then are
\[
v_1(\xi) = \frac{1}{\sqrt{1+\xi_2^2\, [\frac{q'}{q}(\xi_1)]^2}}\, , \qquad
v_2(\xi) = \frac{\xi_2\, \frac{q'}{q}(\xi_1)}{\sqrt{1+\xi_2^2\, [\frac{q'}{q}(\xi_1)]^2}}\, ,
\]
and the Jacobian of $v$ is
\[
v'(\xi) = \left(1+ \left(\xi_2\, \frac{q'}{q}\right)^2\right)^{-3/2} 
\left( \begin{array}{cc} -\xi_2^2\, \frac{q'}{q} \left(\frac{q'}{q}\right)' & -\xi_2 \left(\frac{q'}{q}\right)^2 \\ \xi_2 \left(\frac{q'}{q}\right)' & \frac{q'}{q} \end{array} \right),
\]
with $q, q', q''$ evaluated at $\xi_1$. 
Note that if $\theta$ varies in an interval of the form $[-b,b]$ $(b>0)$, the context curves lie symmetrically about the $x_1$ axis; they are symmetrical \wrt~the $x_2$ axis if $q$ is an even function. The four contexts C1 to C4 used in the experiment were specified by the functions $q \equiv \{q(u), \ |u| \le 2r\}$ given in the following table; $r$ is the radius of the respective target circle. The resulting figures were scaled such that the target circles had diameter 9~cm in the physical stimuli presented to the observers.

\medskip
\begin{tabular}{ccc}
context &function $q(u)$& radius $r$\\
\hline
C1 & $1+u^2$		& 0.5\\
C2 & $1+1/\sqrt{1+u^2}$	& 2.0\\
C3 & $(u+3)^2$		& 1.0\\
C4 & $1$		& 1.0
\end{tabular}

\medskip
\noindent
Supplementary material about experimental details is available from the authors upon request.
%\PDF\ files used for stimulus presentation with contexts C1 to C4 are available as supplementary online material.

% ----------------------------------------------------------------------
\small

%\vspace{-2ex}

\end{document}